\documentclass[aoas,nameyear,MSNbibl,dvips]{arximspdf}
\usepackage{dcolumn}
\usepackage{graphicx}

\doi{10.1214/10-AOAS370}
\volume{5}
\issue{1}
\pubyear{2011}
\firstpage{578}
\lastpage{603}

\makeatletter

\newcolumntype{d}[1]{D{.}{.}{#1}}
\newproclaim{remark}{Remark}
\newproclaim{example}{Example}
\newcommand{\norm}[1]{\Vert#1\Vert}

\newcommand{\Real}{\mathbb{R}}
\newcommand{\To}{\longrightarrow}
\newcommand{\Exp}{\operatorname{Exp}}
\newcommand{\Log}{\operatorname{Log}}
\def\argmin{\mathop{\mathrm{argmin}}}
\makeatother

\begin{document}
\begin{frontmatter}

\title{Principal arc analysis on direct product manifolds}
\runtitle{Principal arc analysis}

\begin{aug}
\author[A]{\fnms{Sungkyu} \snm{Jung}\corref{}\ead[label=e1]{sungkyu@email.unc.edu}},
\author[B]{\fnms{Mark} \snm{Foskey}\ead[label=e2]{mark\_foskey@unc.edu}}
\and
\author[B]{\fnms{J. S.} \snm{Marron}\ead[label=e3]{marron@email.unc.edu}}

\runauthor{S. Jung, M. Foskey and J. S. Marron}

\affiliation{University of North Carolina,
University of North Carolina and\break
University of North Carolina}

\address[A]{S. Jung\\
J. S. Marron\\
Department of Statistics \\
and Operations Research \\
University of North Carolina \\
Chapel Hill, North Carolina 27599 \\
USA\\
\printead{e1}\\
\phantom{E-mail: }\printead*{e3}}

\address[B]{M. Foskey\\
Department of Radiation Oncology\\
University of North Carolina \\
Chapel Hill, North Carolina 27599 \\
USA\\
\printead{e2}}

\end{aug}

\received{\smonth{12} \syear{2009}}
\revised{\smonth{5} \syear{2010}}

%
\begin{abstract}
We propose a new approach to analyze data that naturally lie on
manifolds. We focus on a special class of manifolds, called direct
product manifolds, whose intrinsic dimension could be very high. Our
method finds a low-dimensional representation of the manifold that can
be used to find and visualize the principal modes of variation of the
data, as Principal Component Analysis (PCA) does in linear spaces.
The proposed method improves upon earlier manifold extensions of PCA by
more concisely capturing important nonlinear modes. For the special
case of data on a sphere, variation following nongeodesic arcs is
captured in a single mode, compared to the two modes needed by previous
methods. Several computational and statistical challenges are resolved.
The development on spheres forms the basis of principal arc analysis on
more complicated manifolds.
The benefits of the method are illustrated by a data example using
medial representations in image analysis.
\end{abstract}

%

\end{frontmatter}

\section{Introduction}\label{sec:intro}

Principal Component Analysis (PCA) has been frequently used as a method
of dimension reduction and data visualization for high-dimensional
data. For data that naturally lie in a curved manifold, application of
PCA is not straightforward since the sample space is not linear.
Nevertheless, the need for PCA-like methods is growing as more manifold
data sets are encountered and as the dimensions of the manifolds increase.

In this article we introduce a new approach for an extension of PCA on
a special class of manifold data. We focus on direct products of simple
manifolds, in particular, of the unit circle $S^1$, the unit sphere
$S^2$, $\Real_+$ and $\Real^p$. We will refer to these as \textit{direct
product manifolds}, for convenience.
Many types of statistical sample spaces are special cases of the direct
product manifold. A~widely known example is the sample space for
directional data [\citet{Fisher1993}, \citet{Fisher1993a} and \citet{Mardia2000}] and their direct products. Applications include analysis
of wind directions, orientations of cracks, magnetic field directions
and directions from the earth to celestial objects. For example, when
we consider multiple 3-D directions simultaneously, the sample space is
$S^2 \otimes\cdots\otimes S^2$, which is a direct product manifold.
Another example is the medial representation of shapes [m-reps, \citet{Siddiqi2008}], which is somewhat less known to the statistical
community but provides a powerful parametrization of 3-D shapes of
human organs and has been extensively studied in the image analysis
field. The space of m-reps is usually a high-dimensional direct product
manifold. Some background and necessary definitions on direct product
manifolds can be found in \hyperref[append:manifold]{Appendix}.

Our approach to a manifold version of PCA builds upon earlier work,
especially the principal geodesic analysis proposed by \citet{Fletcher2004a} and the geodesic PCA proposed by \citet{Huckemann2006}
and \citet{Huckemann2009}. A~detailed catalogue of current methodologies
can be found in \citet{Huckemann2009}.
An important approach among these is to approximate the manifold by a
linear space. \citet{Fletcher2004a} take the tangent space of the
manifold at the geodesic mean as the linear space, and work with
appropriate mappings between the manifold and the tangent space. This
results in finding the best fitting geodesics among those passing
through the geodesic mean. This was improved in an important way by
Huckemann, Hotz and Munk, who found the best fit over the set of all
geodesics. Huckemann, Hotz and Munk went on to propose a new notion of
center point, the \textit{PCmean}, which is an intersection of the
first two principal geodesics. This approach gives significant
advantages, especially when the curvature of the manifold makes the
geodesic mean inadequate, an example of which is depicted in Figure~\ref{fig:smallCircleExamples}b.

Our method inherits advantages of these methods and improves further by
effectively capturing more complex nongeodesic modes of variation.
Note that the curvature of direct product manifolds is mainly due to
the spherical part, which motivates careful investigation of
$S^2$-valued variables.
We point out that (small) circles in $S^2$, including geodesics, can be
used to capture the nongeodesic variation.
We introduce the \textit{principal circles} and \textit{principal
circle mean}, analogous to, yet more flexible than, the geodesic
principal component and PCmean of Huckemann, Hotz and Munk. These become
\textit{principal arcs} when the manifold is indeed $S^2$. For more
complex direct product manifolds, we suggest transforming the data
points in $S^2$ into a linear space by a special mapping utilizing the
principal circles. For the other components of the manifold, the
tangent space mappings can be used to map the data into a linear space
as done in \citet{Fletcher2004a}.
Once manifold-valued data are mapped onto the linear space, then the
classical linear PCA can be applied to find principal components in the
transformed linear space.
The estimated principal components in the linear space can be
back-transformed to the manifold, which leads to principal arcs.

\begin{figure}

\includegraphics{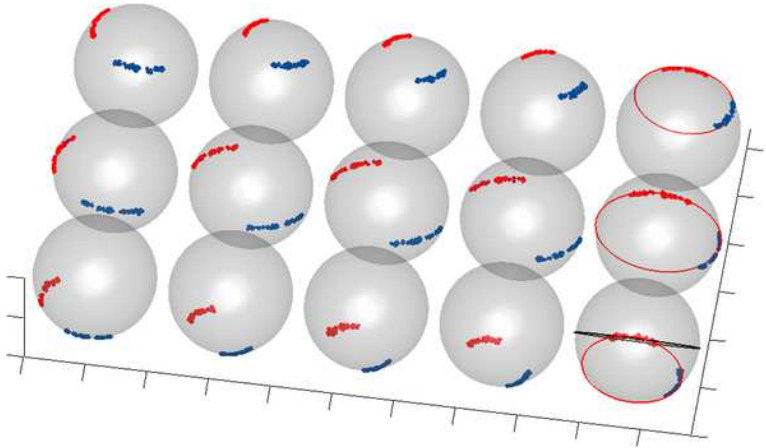}

\caption{$S^2$-valued samples ($n = 60$) of the prostate m-reps with 15
medial atoms. One sample in this figure is represented by a 30-tuple of
3-D directions (two directions at each atom), which lies in the
manifold $\bigotimes_{i=1}^{15} (S^2 \otimes S^2)$. Small and great
circles are fitted and plotted in the rightmost atoms to emphasize the
sample variation along small circles. }
\label{fig:prostateExample}
\end{figure}

We illustrate the potential of our method by an example of m-rep data
in Figure~\ref{fig:prostateExample}.
Here, m-reps with 15 sample points called atoms model the prostate
gland (an organ in the male reproductive system) and come from the
simulator developed and analyzed in \citet{Jeong2008}.
Figure~\ref{fig:prostateExample} shows that the $S^2$ components of the
data tend to be distributed along small circles, which frequently are
not geodesics. We emphasize the curvature of variation along each
sphere by fitting a great circle and a small circle (by the method
discussed in Section~\ref{PAAonSphere}). Our method is adapted to
capture this nonlinear (nongeodesic) variation of the data. A~potential
application of our method is to improve accuracy of segmentation of
objects from CT images. Detailed description of the data and results of
our analysis can be found in Section~\ref{sec:realData}.

Note that the previous approaches [\citet{Fletcher2004a}, \citet{Huckemann2006}] are defined for general manifolds, while our method
focuses on these particular direct product manifolds. Although the
method is not applicable for general manifolds, it is useful for this
common class of manifolds that is often found in applications. Our
results inform our belief that focusing on specific types of manifolds
allow more precise and informative statistical modeling than methods
that attempt to be fully universal. This happens through using special
properties (e.g., presence of small circles) that are not available for
all other manifolds.

The rest of the article is organized as follows.
We begin by introducing a \textit{circle class} on $S^2$ as an
alternative to the set of geodesics. Section~\ref{PAAonSphere}
discusses principal circles in $S^2$, which will be the basis of the
special transformation. The first principal circle is defined by the
least-squares circle, minimizing the sum of squared residuals. In
Section~\ref{sec:smallVSgreat} we introduce a data-driven method to
decide whether the least-squares circle is appropriate.
A~recipe for principal arc analysis on direct product manifolds is
proposed in Section~\ref{PAAonManifold} with discussion on the
transformations. A~detailed introduction of the space of m-reps and the
results from applying the proposed method follow. A~novel computational
algorithm for the least-squares circles is presented in Section~\ref{sec:algorithm}.
In the \hyperref[append:manifold]{Appendix} we provide some necessary background
for treating direct product manifolds as sample spaces, including the
notion of geodesic mean, tangent space, exponential map and log map.

\section{Circle class for nongeodesic variation on $S^2$}\label{PAAonSphere}

Consider a set of points in $\Real^2$. Numerous methods for
understanding population properties of a data set in linear space have
been proposed and successfully applied, which include rigid methods,
such as linear regression and principal components, and very flexible
methods, such as scatterplot smoothing and principal curves [\citet{Hastie1989}]. In this paper we make use of a parametric class of
circles, including small and great circles, which allows much more
flexibility than either methods of \citet{Fletcher2004a} or \citet{Huckemann2009}, but less flexibility than a principal curve approach.
Although this idea was motivated by examples such as those in
Figure~\ref{fig:prostateExample}, there are more advantages gained from
using the class of circles:
\begin{enumerate}[(iii)]
\item[(i)] The circle class includes the simple geodesic case.
\item[(ii)] Each circle can be parameterized, which leads to an easy
interpretation.
\item[(iii)] There is an orthogonal complement of each circle, which
gives two important advantages:
\begin{enumerate}[(a)]
\item[(a)] Two orthogonal circles can be used as a basis of a further
extension to principal arc analysis.
\item[(b)] Building a sensible notion of principal components on $S^2$
alone is easily done by utilizing the circles.
\end{enumerate}
\end{enumerate}

The idea (iii)(b) will be discussed in detail after introducing a method
of circle fitting. Some notations follow: $S^2$ can be thought of as
the unit sphere in $\Real^3$, so that a unit vector $\mathbf{x} \in
\Real^3$ is a member of $S^2$. The geodesic distance between $\mathbf{x}, \mathbf{y} \in S^2$, denoted by $\rho( \mathbf{x}, \mathbf{y})$,
is defined as the shortest distance between $\mathbf{x}$, $\mathbf{y}$
along the sphere, which is the same as the angle formed by the two
vectors. Thus, $\rho( \mathbf{x},\mathbf{y}) = \cos^{-1}(\mathbf{x}'\mathbf{y})$. A~circle on $S^2$ is conveniently parameterized by
center $\mathbf{c} \in S^2$ and geodesic radius $r$, and denoted by
$\delta(\mathbf{c},r) = \{\mathbf{x} \in S^2 | \rho(\mathbf{c},\mathbf{x}) = r \}$. It is a geodesic when $r = \pi/2$. Otherwise it is a
small circle.

A circle that best fits the points $\mathbf{x}_1,\ldots, \mathbf{x}_n
\in S^2$ is found by minimizing the sum of squared residuals. The
residual of $\mathbf{x}_i$ is defined as the signed geodesic distance
from $\mathbf{x}_i$ to the circle $\delta(\mathbf{c},r)$. Then the
least-squares circle is obtained by
%
\begin{eqnarray}\label{eq:getSmallCircle}
\min_{\mathbf{c} , r} \sum_{i=1}^{n}  \bigl(\rho(\mathbf{x}_i, \mathbf{c}) - r  \bigr)^2 \qquad
\mbox{subject to } \mathbf{c}\in S^2, r \in(0, \pi).
\end{eqnarray}
Note that there are always multiple solutions of (\ref{eq:getSmallCircle}). In particular, whenever $(\mathbf{c},r)$ is a
solution, $(-\mathbf{c},\pi- r)$ also solves the problem as $\delta
(\mathbf{c},r) = \delta(-\mathbf{c},\pi- r)$. This ambiguity does not
affect any essential result in this paper. Our convention is to use the
circle with smaller geodesic radius.

The optimization task (\ref{eq:getSmallCircle}) is a constrained
nonlinear least squares problem. We propose an algorithm to solve the
problem that features a simplified optimization task and approximation
of $S^2$ by tangent planes. The algorithm works in a doubly iterative
fashion, which has been shown by experience to be stable and fast.
Section~\ref{sec:algorithm} contains a detailed illustration of the algorithm.

Analogous to principal geodesics in $S^2$, we can define \textit{principal circles} in $S^2$ by utilizing the least-squares circle. The
principal circles are two orthogonal circles in $S^2$ that best fit the
data. We require the first principal circle to minimize the variance of
the residuals, so it is the least-squares circle (\ref{eq:getSmallCircle}). The second principal circle is a geodesic which
passes through the center of the first circle and thus is orthogonal at
the points of intersection. Moreover, the second principal circle is
chosen so that one intersection point is the intrinsic mean [defined in
(\ref{eq:princCircMean}) later] of the projections of the data onto the
first principal circle.

\begin{figure}
\centering%
{\begin{tabular}{@{}cc@{}}

\includegraphics{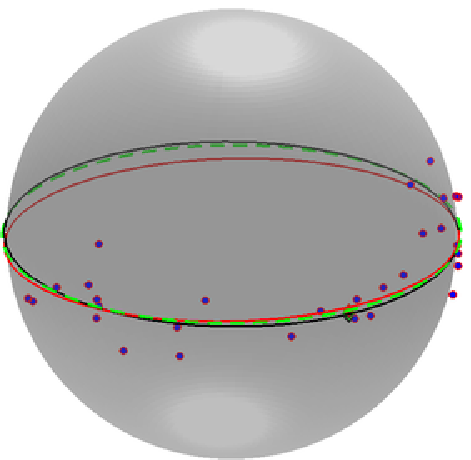}
 & \includegraphics{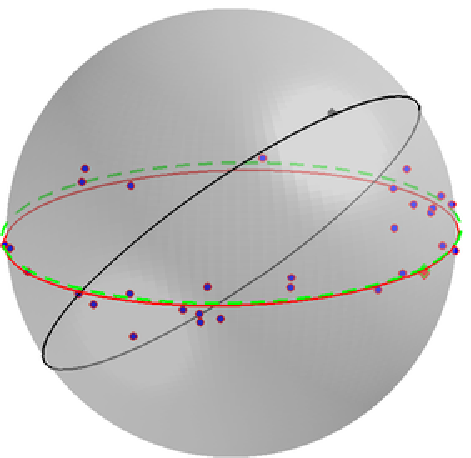}\\
(a) & (b)\\[6pt]

\includegraphics{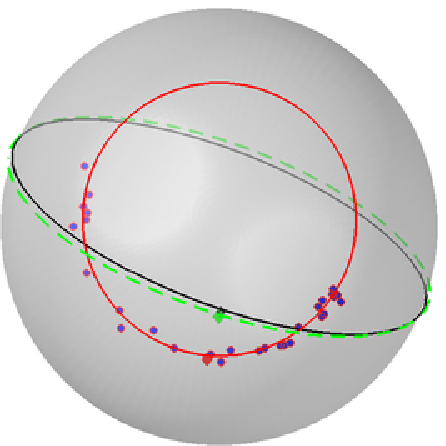}
 & \includegraphics{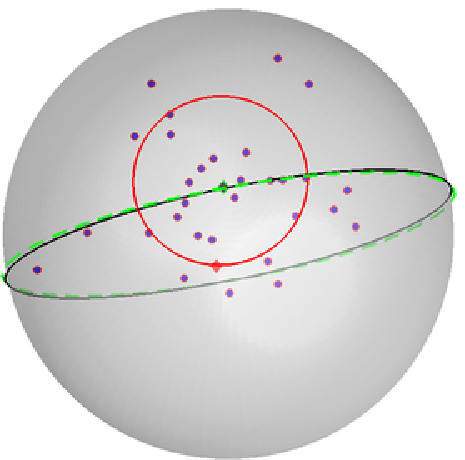}\\
(c) & (d)
\end{tabular}}
\caption{Toy examples on $S^2$ with $n = 30$ points showing the first
principal circle (red) as a small circle and the first geodesic
principal component (dotted green) by Huckemann, and the first
principal geodesic (black) by Fletcher. Also plotted are the geodesic
mean, PCmean and principal circle mean of the data as black, green and
red diamonds, respectively.
\textup{(a)} The three methods give similar satisfactory answers when the data
are stretched along a geodesic.
\textup{(b)} When the data are stretched along a great circle, covering almost
all of it, the principal geodesic (black circle) and geodesic mean
(black diamond) fail to find a reasonable representation of the data,
while the principal circle and Huckemann's geodesic give sensible answers.
\textup{(c)} Only the principal circle fits well when the data are not along a geodesic.
\textup{(d)} For a small cluster without principal modes of variation, the
principal circle gets too small. See Section~\protect\ref{sec:smallVSgreat} for
discussion of this phenomenon.
}
\label{fig:smallCircleExamples}
\end{figure}

Based on a belief that the intrinsic (or extrinsic) mean defined on a
curved manifold may not be a useful notion of the center point of the
data [see, e.g., \citeauthor{Huckemann2009} and Figure~\ref{fig:smallCircleExamples}b], the principal circles do not use the
pre-determined means. To develop a better notion of center point, we
locate the best 0-dimensional representation of the data in a
data-driven manner. Inspired by the PCmean idea of
\citeauthor{Huckemann2009},
given the first principal circle $\delta_1$, the \textit{principal circle mean} $\mathbf{u} \in\delta_1$ is defined (in an
intrinsic way) as
%
\begin{equation}\label{eq:princCircMean}
\mathbf{u} =\argmin_{\mathbf{u} \in\delta_1} \sum_{i=1}^{n} \rho
^2(\mathbf{u},P_{\delta_1} \mathbf{x}_i),
\end{equation}
where $P_{\delta_1} \mathbf{x}$ is the projection of $\mathbf{x}$ onto
$\delta_1$, that is, the point on $\delta_1$ of the shortest geodesic
distance to $\mathbf{x}$. Then
%
\begin{equation}\label{eq:projectiondelta1}
P_{\delta_1(\mathbf{c},r)} \mathbf{x} = \frac{ \mathbf{x} \sin(r)+
\mathbf{c} \sin(\rho(\mathbf{x},\mathbf{c}) -r)}
{\sin(\rho(\mathbf{x},\mathbf{c}))},
\end{equation}
as in equation (3.3) of \citet{Mardia1977}. We assume that $\mathbf{c}$
is the north pole $\mathbf{e}_3$, without losing generality since
otherwise the sphere can be rotated. Then
%
\begin{equation}\label{eq:intrinsinU}
\rho(\mathbf{u},P_{\delta_1} \mathbf{x}) =
\sin(r) \rho_{S^1}  \biggl( \frac{(u_1,u_2)}{\sqrt{1-u_3^2}}, \frac
{(x_1,x_2)}{\sqrt{1-x_3^2}} \biggr),
\end{equation}
where $\mathbf{u} = (u_1,u_2,u_3)'$, $\mathbf{x} = (x_1,x_2,x_3)'$ and
$\rho_{S^1}$ is the geodesic (angular) distance function on $S^1$. The
optimization problem (\ref{eq:princCircMean}) is equivalent to finding
the geodesic mean in $S^1$. See equation (\ref{eq:geodmeanS1}) in the
\hyperref[append:manifold]{Appendix} for computation of the geodesic mean in $S^1$.

The second principal circle $\delta_2$ is then the geodesic passing
through the principal circle mean $\mathbf{u}$ and the center $\mathbf{c}$ of $\delta_1$. Denote $\bar\delta \equiv\bar\delta(\mathbf{x}_1,\ldots,\mathbf{x}_n)$ as a combined representation of $(\delta_1,
\mathbf{u})$ or, equivalently, $(\delta_1, \delta_2)$.

As a special case, we can force the principal circles to be great
circles. The best fitting geodesic is obtained as a solution of the
problem (\ref{eq:getSmallCircle}) with $r = \pi/ 2$ and becomes the
first principal circle. The optimization algorithm for this case is
slightly modified from the original algorithm for the varying $r$ case
by simply setting $r = \pi/ 2$. The principal circle mean $\mathbf{u}$
and the $\delta_2$ for this case are defined in the same way as in the
small circle case. Note that the principal circles with $r = \pi/2$ are
essentially the same as the method of \citet{Huckemann2006}.

Figure~\ref{fig:smallCircleExamples} illustrates the advantages of
using the circle class to efficiently summarize variation. On four
different sets of toy data, the first principal circle $\delta_1$ is
plotted with principal circle mean $\mathbf{u}$. The first principal
geodesics from the methods of Fletcher and Huckemann are also plotted
with their corresponding mean.
Figure~\ref{fig:smallCircleExamples}a illustrates the case where the data
were indeed stretched along a geodesic. The solutions from the three
methods are similar to one another.
The advantage of Huckemann's method over Fletcher's can be found in
Figure~\ref{fig:smallCircleExamples}b. The geodesic mean is found far from
the data, which leads to poor performance of the principal geodesic
analysis, because it considers only great circles passing through the
geodesic mean. Meanwhile, the principal circle and Huckemann's method,
which do not utilize the geodesic mean, work well.
The case where geodesic mean and any geodesic do not fit the data well
is illustrated in Figure~\ref{fig:smallCircleExamples}c, which is analogous
to the Euclidean case, where a nonlinear fitting may do a better job of
capturing the variation than PCA. To this data set, the principal
circle fits best, and our definition of mean is more sensible than the
geodesic mean and the PCmean.
The points in Figure~\ref{fig:smallCircleExamples}d are generated from the
von Mises--Fisher distribution with $\kappa= 10$, thus having no
principal mode of variation. In this case the first principal circle
$\delta_1$ follows a contour of the apparent density of the points. We
shall discuss this phenomenon in detail in the following section.

Fitting a (small) circle to data on a sphere has been investigated for
some time, especially in statistical applications in geology. Those
approaches can be distinguished in three different ways, where our
choice fits into the first category:
\begin{enumerate}[(2)]
\item[(1)] Least-squares of intrinsic residuals: \citet{Gray1980} formulated
the same problem as in (\ref{eq:getSmallCircle}), finding a circle that
minimizes sum of squared residuals, where residuals are defined in a
geodesic sense.
\item[(2)] Least-squares of extrinsic residuals: A different measure of
residual was chosen by \citet{Mardia1977} and \citet{Rivest1999}, where
the residual of $\mathbf{x}$ from $\delta(c,r)$ is defined by the
shortest Euclidean distance between $\mathbf{x}$ and $\delta(\mathbf{c},r)$. Their objective is to find
\[
\argmin_\delta\sum_{i=1}^n \norm{\mathbf{x}_i - P_\delta\mathbf{x}_i}^2
= \argmin_\delta\sum_{i=1}^n -\mathbf{x}_i' P_\delta\mathbf{x}_i
= \argmin_\delta\sum_{i=1}^n -\cos(\xi_i),
\]
where $\xi_i$ denotes the intrinsic residual. This type of approach can
be numerically close to the intrinsic method as $\cos(\xi_i) = 1- \xi
_i^2 / 2 + O(\xi_i^4)$.
\item[(3)] Distributional approach: \citet{Mardia1977} and \citet{Bingham1978}
proposed appropriate distributions to model $S^2$-valued data that
cluster near a small circle. These models essentially depend on the
quantity $\cos(\xi)$, which is easily interpreted in the extrinsic
sense but not in the intrinsic sense.
\end{enumerate}

\begin{remark*}
The principal circle and principal circle mean always exist. This is
because the objective function (\ref{eq:getSmallCircle}) is a
continuous function of $\mathbf{c}$, with the compact domain $S^2$. The
minimizer $r$ has a closed-form solution (see Section~\ref{sec:algorithm}). A~similar argument can be made for the existence of
$\mathbf{u}$. On the other hand, the uniqueness of the solution is not
guaranteed. We conjecture that if the manifold is approximately linear
or, equivalently, the data set is well approximated by a linear space,
then the principal circle will be unique. However, this does not lead
to the uniqueness of $\mathbf{u}$, whose sufficient condition is that
the projected data on $\delta_1$ is strictly contained in a half-circle
[\citet{Karcher1977}]. Note that a sufficient condition for the
uniqueness of the principal circle is not clear even in the Euclidean
case [\citet{Chernov2010}].
\end{remark*}

\section{Suppressing small least-squares circles}\label{sec:smallVSgreat}

When the first principal circle $\delta_1$ has a small radius,
sometimes it is observed that $\delta_1$ does not fit the data in a
manner that gives useful decomposition, as shown in Figure~\ref{fig:smallCircleExamples}d. This phenomenon has been also observed for the
related principal curve fitting method of \citet{Hastie1989}. We view
this as unwanted overfitting, which is indeed a side effect caused by
using the full class of circles with free radius parameter instead a
class of great circles. In this section a data-driven method to flag
this overfitting is discussed. In essence, the fitted small circle is
replaced by the best fitting geodesics when the data do not cluster
along the circle but instead tend to cluster near the center of the circle.

We first formulate the problem and solution in $\Real^2$. This is for
the sake of clear presentation and also because the result on $\Real^2$
can be easily extended to $S^2$ using a tangent plane approximation.

Let $f_\mathbf{X}$ be a spherically symmetric density function of a
continuous distribution defined on $\Real^2$. Whether the density is
high along some circle is of interest. By the symmetry assumption,
density height along a circle can be found by inspecting a section of
$f_\mathbf{X}$ along a ray from the origin (the point of symmetry).
A~section of $f_\mathbf{X}$ coincides with the conditional density
$f_{X_1 |X_2} (x_1 | x_2 = 0) = \kappa^{-1} f_\mathbf{X}(x_1,0)$.
A~random variable corresponding to the p.d.f. $f_{X_1 |X_2=0}$ is not
directly observable. Instead, the radial distance $R = \norm{\mathbf{X}}$ from the origin can be observed.
For the polar coordinates $(R, \Theta)$ such that $\mathbf{X} =
(X_1,X_2) = (R\cos\Theta, R\sin\Theta)$, the marginal p.d.f. of $R$ is
$f_R(r) = 2\pi r f_\mathbf{X}(r,0)$ as $f_\mathbf{X}$ is spherically
symmetric. A~section of $f_\mathbf{X}$ is related to the observable
density $f_R$ as $f_R(r) \propto r f_{X_1 |X_2=0}(r)$, for $r\ge0$.
This relation is called the length-biased sampling problem [\citet{Cox1969}]. The relation can be understood intuitively by observing that
a value $r$ of $R$ can be observed at any point on a circle of radius
$r$, circumference of which is proportional to $r$. Thus, sampling of
$R$ from the density $f_{X_1 |X_2=0}$ is proportional to its size.

The problem of suppressing a small circle can be paraphrased as ``how
to determine whether a nonzero point is a mode of the function $f_{X_1
|X_2=0}$, when observing only a length-biased sample.''

\begin{figure}

\includegraphics{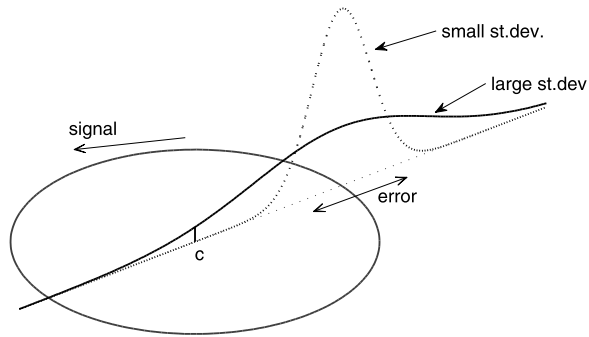}

\caption{Illustration of the conceptual model (\protect\ref{eq:modelForError})
on $\Real^2$, which can also be understood as a local approximation of
$S^2$. The signal is along the circle centered at $c$ and radius $\mu$.
The error is perpendicular to the signal. When the deviation $\sigma$
is large, it is possible that the amount of error is even greater than
the radius $\mu$. This is incorporated in the wrapping approach (\protect\ref{eq:wrapping}).}
\label{fig:scDist}
\end{figure}

The spectrum from the circle-clustered case (mode at a nonzero point)
to the center-clustered case (mode at origin) can be modeled as
%
\begin{equation}\label{eq:modelForError}
\mathrm{data} = \mathrm{signal} + \mathrm{error},
\end{equation}
where the signal is along a circle with radius $\mu$, and the error
accounts for the perpendicular deviation from the circle (see
Figure~\ref{fig:scDist}). Then, in polar coordinates $(R,\Theta)$,
$\Theta$ is uniformly distributed on $(0,2\pi]$ and $R$ is a positive
random variable with mean $\mu$. First assume that $R$ follows a
truncated Normal distribution with standard deviation $\sigma$, with
the marginal p.d.f. proportional to
%
\begin{equation}\label{eq:f_R}
f_R(r) \propto\phi \biggl( \frac{r-\mu}{\sigma} \biggr)\qquad \mbox{for }r\ge0,
\end{equation}
where $\phi$ is the standard Normal density function. The conditional
density $f_{X_1|X_2 = 0}$ is then
\[
f_{X_1|X_2 = 0}(r) \propto\frac{1}{r}f_R(r) \propto\frac{1}{r\sigma}
\exp \biggl(-\frac{(r-\mu)^2}{\sigma^2} \biggr)\qquad \mbox{for }r > 0.
\]
Nonzero local extrema of $f_{X_1|X_2 = 0}$ can be characterized as a
function of $(\mu, \sigma)$ in terms of $r_+, r_- = \{ \mu\pm\sqrt
{(\mu-2\sigma)(\mu+2\sigma)} \} / 2$ as follows:
\begin{itemize}
\item When $\mu> 2\sigma$, $f_{X_1|X_2 = 0}$ has a local maximum at
$r_+$, minimum at $r_-$.
\item When $\mu= 2\sigma$, $r_+ = r_- = \frac{\mu}{2}.$
\item When $\mu< 2\sigma$, $f_{X_1|X_2 = 0}$ is strictly decreasing,
for $r\ge0$.
\end{itemize}
Therefore,\vspace*{1pt} whenever the ratio $\mu/ \sigma>2$, $f_{X_1|X_2 = 0}$ has a
mode at $r_+$.

\begin{figure}

\includegraphics{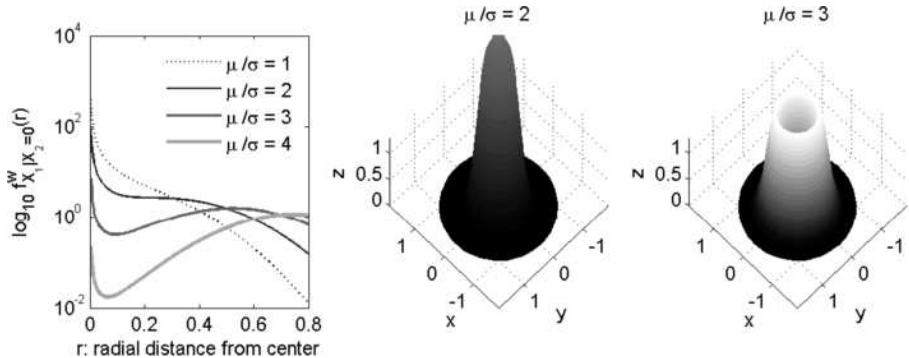}

\caption{(Left)\vspace*{1pt} Graph of $f_{X_1|X_2=0}^w(r)$ for $\mu/\sigma=
1,2,3,4$. The density is high at a nonzero point when $\mu/\sigma>
2.0534$. (Center, right) Spherically symmetric distributions $f$
corresponding to $\mu/\sigma= 2,3$. The ratio $\mu/\sigma> 2$ roughly
leads to a high density along a circle.}
\label{fig:DensityRadialSectSurface}
\end{figure}

This idea can be applied for circles in $S^2$ with some modification,
shown next. We point out that the model (\ref{eq:modelForError}) is
useful for understanding the small circle fitting: signal as a circle
with radius $\mu$, and error as the deviation along geodesics
perpendicular to the circle. Moreover, a spherically symmetric
distribution centered at $\mathbf{c}$ on $S^2$ can be mapped to a
spherically symmetric distribution on the tangent space at $\mathbf{c}$, preserving the radial distances by the log map (defined in the
\hyperref[append:manifold]{Appendix}). A~modification needs to be made on the truncated density
$f_R$. It is more natural to let the error be so large that the
deviation from the great circle is greater than $\mu$. Then the
observed value may be found near the opposite side of the true signal,
which is illustrated in Figure~\ref{fig:scDist} as the large deviation
case. To incorporate this case, we consider a wrapping approach. The
distribution of errors (on the real line) is \textit{wrapped} around
the sphere along a great circle through $\mathbf{c}$, and the marginal
p.d.f. $f_R$ in (\ref{eq:f_R}) is modified to
%
\begin{equation}\label{eq:wrapping}
\hspace*{20pt}f_R^w (r) \propto\sum_{k=0}^{\infty}  \biggl[
\phi \biggl(\frac{r+2\pi k - \mu}{\sigma} \biggr)+\phi \biggl(\frac{r-2\pi k
+ \mu}{\sigma} \biggr)
 \biggr]\qquad \mbox{for } r \in[0,\pi].
\end{equation}
The corresponding conditional p.d.f., $f_{X_1|X_2=0}^w$, is similar to
$f_{X_1|X_2=0}$ and a numerical calculation shows that
$f_{X_1|X_2=0}^w$ has a mode at some nonzero point whenever $\mu/\sigma
> 2.0534$, for $\mu< \pi/2$. In other words, we use the small circle
when $\mu/\sigma$ is large. Note that in what follows we only consider
the first term ($k=0$) of (\ref{eq:wrapping}) since other terms are
negligible in most situations. We have plotted $f_{X_1|X_2=0}^w$ for
some selected values of $\mu$ and $\sigma$ in Figure~\ref{fig:DensityRadialSectSurface}.

With a data set on $S^2$, we need to estimate $\mu$ and $\sigma$, or
the ratio $\mu/\sigma$. Let $\mathbf{x}_1,\ldots, \mathbf{x}_n \in S^2$
and let $\hat{\mathbf{c}}$ be the samples and the center of the fitted
circle, respectively.
Denote $\xi_i$ for the errors of the model (\ref{eq:modelForError})
such that $\xi_i \sim N(0,\sigma^2)$. Then $r_i \equiv\rho(\mathbf{x}_i,\hat{\mathbf{c}}) = | \mu+\xi_i |$, which has the \textit{folded
normal} distribution [\citet{Leone1961}]. Estimation of $\mu$ and $\sigma
$ based on unsigned $r_i$ is not straightforward. We present two
different approaches to this problem.

\textit{Robust approach}.
The observations $r_1, \ldots, r_n$ can be thought of as a set of
positive numbers contaminated by the folded negative numbers.
Therefore, the left half (near zero) of the data are more contaminated
than the right half. We only use the right half of the data, which are
less contaminated than the other half.
We propose to estimate $\mu$ and $\sigma$ by
%
\begin{equation}\label{eq:robustEstimation}
\hat\mu= \operatorname{med}(r_1^n),\qquad
\hat\sigma= \bigl(\mbox{Q}_3 (r_1^n) -\operatorname{med}(r_1^n)\bigr)/\mbox{Q}_3 (\Phi),
\end{equation}
where $\mbox{Q}_3 (\Phi)$ is the third quantile of the standard normal
distribution. The ratio can be estimated by $\hat\mu/ \hat\sigma$.

\textit{Likelihood approach via EM algorithm}.
The problem may also be solved by a likelihood approach. Early
solutions can be found in Leone, Nelson and Nottingham (\citeyear{Leone1961}),
\citet{Elandt1961} and
\citet{Johnson1962}, in which the MLEs were given by numerically solving
nonlinear equations based on the sample moments. As those methods were
very complicated, we present a simpler approach based on the EM
algorithm. Consider unobserved binary variables $s_i$ with values $-1$
and $+1$ so that $s_i r_i \sim N(\mu, \sigma^2)$. The idea of the EM
algorithm is that if we have observed $s_i$, then the maximum
likelihood estimator of $\vartheta=(\mu, \sigma^2)$ would be easily
obtained. The EM algorithm is an iterative algorithm consisting of two
steps. Suppose that the $k$th iteration produced an estimate $\hat
\vartheta_k$ of $\vartheta$.
The E-step is to impute $s_i$ based on $r_i$ and $\hat\vartheta_k$ by
forming a conditional expectation of log-likelihood for $\vartheta$,
\begin{eqnarray*}
Q(\vartheta)
&=& E  \Biggl[ \log\prod_{i=1}^{n} f ( r_i, s_i | \vartheta)  \Big| r_i, \hat\vartheta_k  \Biggr]\\
&=& \sum_{i=1}^{n}  \bigl[\log f(r_i | s_i = +1 , \vartheta) P (s_i = +1
| r_i , \hat\vartheta_k )
\\
&&\hspace*{17pt}{}+ \log f(r_i | s_i = -1 , \vartheta) P (s_i = -1 | r_i , \hat\vartheta_k ) \bigr]\\
&=& \sum_{i=1}^{n}  \bigl[\log\phi(r_i |\vartheta) p_{i(k)}
+ \log\phi(-r_i | \vartheta) \bigl(1-p_{i(k)}\bigr)  \bigr],
\end{eqnarray*}
where $f$ is understood as an appropriate density function, and
$p_{i(k)}$ is easily computed as
\begin{eqnarray*}
p_{i(k)} =
P (s_i = +1 | r_i , \hat\vartheta_k )
= \frac{\phi(r_i|\hat\vartheta_k)}
{\phi(r_i|\hat\vartheta_k)+
\phi(-r_i|\hat\vartheta_k)}.
\end{eqnarray*}
The M-step is to maximize $Q(\vartheta)$ whose solution becomes the
next estimator $\hat\vartheta_{k+1}$. Now the $(k+1)$th estimates are
calculated by a simple differentiation and given by
\begin{eqnarray*}
\hat\mu_{k+1} = \frac{1}{n}\sum_{i=1}^{n} \bigl(2p_{i(k)} - 1\bigr) r_i,\qquad
\hat\sigma^2_{k+1} = \frac{1}{n}\sum_{i=1}^{n}  (r_i^2 - \hat\mu
_{k+1}^2 ).
\end{eqnarray*}

%

With the sample mean and variance of $r_1,\ldots, r_n$ as an initial
estimator $\hat\vartheta_0$, the algorithm iterates E-steps and M-steps
until the iteration changes the estimates less than a predefined
criteria (e.g., $10^{-10}$). $\mu/\sigma$ is estimated by the ratio of
the solutions.

\begin{table}[b]
\caption{Proportion of estimates greater than $2$ from the data
illustrated in Figure~\protect\ref{fig:Estimators}. For $\mu/\sigma=3$, shown
are proportions of correct answers from each estimator. For $\mu/\sigma
= 1$ or 0, shown are proportions of \textit{incorrect} answers}
\label{tab:Estimators}
\begin{tabular*}{\textwidth}{@{\extracolsep{4in minus 4in}}ld{3.1}d{2.1}d{1.1}d{1.1}@{}}
\hline
\textbf{Method}
& \multicolumn{1}{c}{$\bolds{\mu/\sigma=3}$}
& \multicolumn{1}{c}{$\bolds{\mu/\sigma=2}$}
& \multicolumn{1}{c}{$\bolds{\mu/\sigma=1}$}
& \multicolumn{1}{c@{}}{$\bolds{\mu/\sigma=0}$} \\
\hline
MLE, $n =50$ & 98.5 & 55.2& 5.2 & 6.8 \\
Robust, $n = 50$ & 95.0 & 50.5& 4.7 & 1.4 \\
MLE, $n =1000$ & 100& 51.9& 0 & 0 \\
Robust, $n = 1000$ & 100& 50.5& 0 & 0 \\
\hline
\end{tabular*}
\end{table}

\begin{figure}

\includegraphics{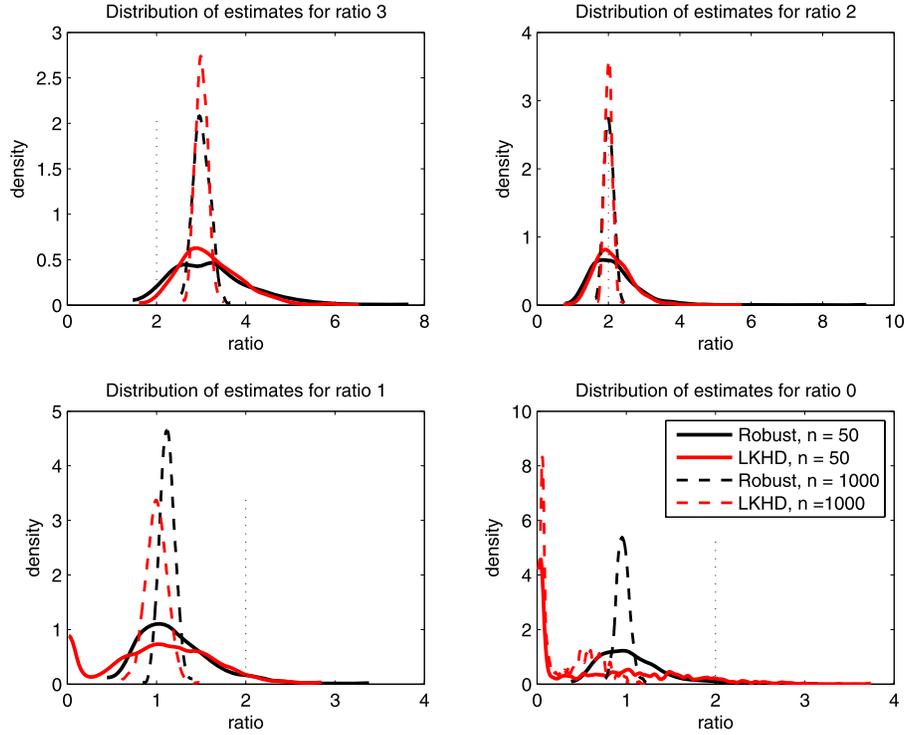}

\caption{Simulation results of the proposed estimators for the ratio
$\mu/\sigma$. Different ratios represent different underlying
distributions. For example, estimators in the top left are based on
random samples from a folded Normal distribution with mean $\mu= 3$,
standard deviation $\sigma= 1$. Curves are smooth histograms of
estimates from 1000 repetitions. The thick black curve represents the
distribution of the robust estimator from $n=50$ samples. Likewise, the
thick red curve is for the MLE with $n=50$, the dotted black curve is
for the robust estimator with $n=1000$, and the dotted red curve is for
the MLE with $n=1000$. The smaller sample size represents a usual data
analytic situation, while the $n=1000$ case shows an asymptotic situation.}
\label{fig:Estimators}
\end{figure}

\textit{Comparison}. Performance of these estimators are now examined by
a simulation study. Normal random samples are generated with ratios $\mu
/\sigma$ being 0, 1, 2 or 3, representing the transition from the
center-clustered to circle-clustered case. For each ratio, $n = 50$
samples are generated, from which $\hat\mu/ \hat\sigma$ is estimated.
These steps are repeated 1000 times to obtain the sampling variation of
the estimates. We also study the $n = 1000$ case in order to
investigate the consistency of the estimators. The results are
summarized in Figure~\ref{fig:Estimators} and Table~\ref{tab:Estimators}.

The distribution of estimators are shown for $n=50, 1000$ in Figure~\ref{fig:Estimators} and the proportion of estimators greater than 2 is
summarized in Table~\ref{tab:Estimators}. When $n = 1000$, both
estimators are good in terms of the proportion of correct answers. In
the following, the proportions of correct answers are corresponding to
$n=50$ case.
The top left panel in Figure~\ref{fig:Estimators} illustrates the
circle-centered case with ratio 3. The estimated ratios from the robust
approach give correct solutions (greater than 2) 95\% of the time
(98.5\% for likelihood approach). For the borderline case (ratio 2, top
right), the small circle will be used about half the time. The
center-clustered case is demonstrated with the true ratio 1, that also
gives a reasonable answer (proportion of correct answers 95.3\% and
94.8\% for the robust and likelihood answers respectively). It can be
observed that when the true ratio is zero, the robust estimates are far
from 0 (the bottom right in Figure~\ref{fig:Estimators}). However, this
is expected to occur because the proportion of uncontaminated data is
low when the ratio is too small. However, those `inaccurate' estimates
are around 1 and less than 2 most of the time, which leads to `correct'
answers. The likelihood approach looks somewhat better with more hits
near zero, but an asymptotic study [\citet{Johnson1962}] showed that the
variance of the maximum likelihood estimator converges to infinity when
the ratio tends to zero, as glimpsed in the long right tail of the
simulated distribution.

In summary, we recommend use of the robust estimators (\ref{eq:robustEstimation}), which are computationally light,
straightforward and stable for all cases.

In addition, we point out that \citet{Gray1980} and \citet{Rivest1999}
proposed to use a goodness-of-fit statistic to test whether the small
circle fit is better than a geodesic fit. Let $r_g$ and $r_c$ be the
sums of squares of the residuals from great and small circle fits. They
claimed that $V = (n-3)(r_g - r_c)/r_c$ is approximately distributed as
$F_{1,n-3}$ for a large $n$ if the great circle was true. However, this
test does not detect the case depicted in Figure~\ref{fig:smallCircleExamples}d. The following numerical example shows the
distinction between our approach and the goodness-of-fit approach.
\begin{example}
Consider the sets of data depicted in Figure~\ref{fig:smallCircleExamples}. The goodness-of-fit test gives $p$-values of
$0.51$, $0.11359$, $0$ and $0.0008$ for (a)--(d), respectively. The
estimated ratios $\mu/\sigma$ are $14.92, 16.89$, $14.52$ and $1.55$.
Note that for (d), when the least-squares circle is too small, our
method suggests to use a geodesic fit over a small circle while the
goodness-of-fit test gives significance of the small circle. The
goodness-of-fit method is not adequate to suppress the overfitting
small circle in a way we desire.
\end{example}

\begin{figure}

\includegraphics{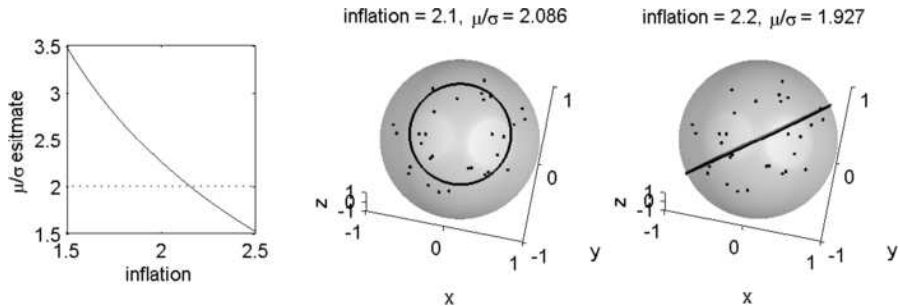}

\caption{(Left) The estimate $\widehat{\mu/\sigma}$ decreases smoothly
as the perturbation is inflated. (Center, right) Snapshots of the toy
data on a sphere. A~very small perturbation of the data set leads to a
sharp transition between the small circle (center) and great circle (right).}
\label{fig:snapping}
\end{figure}

A referee pointed out that the transition of the principal circle
between great circle and small circle is not continuous. Specifically,
when the data set is perturbed so that the principal circle becomes too
small, then the principal circle and principal circle mean are abruptly
replaced by a great circle and geodesic mean. As an example, we have
generated a toy data set spread along a circle with some radial
perturbation. The perturbation is continuously inflated, so that with
large inflation, the data are no longer circle-clustered. In Figure~\ref{fig:snapping} the $\widehat{\mu/\sigma}$ changes smoothly, but once
the estimate hits 2 (our criterion), there is a sharp transition
between small and great circles. Sharp transitions do naturally occur
in the statistics of manifold data. For example, even the simple
geodesic mean can exhibit a major discontinuous transition resulting
from an arbitrarily small perturbation of the data. However, the
discontinuity between small and great circles does seem more arbitrary
and thus may be worth addressing. An interesting open problem is to
develop a blended version of our two solutions, for values of $\widehat
{\mu/\sigma}$ near 2, which could be done by fitting circles with radii
that are smoothly blended between the small circle radius and $\pi/2$.
%

\section{Principal arc analysis on direct product manifolds}\label
{PAAonManifold}

The discussions of the principal circles in $S^2$ play an important
role in defining the principal arcs for data in a direct product
manifold $M = M_1\, \otimes\, M_2\, \otimes\,\cdots\,\otimes\, M_d$, where each
$M_i$ is one of the simple manifolds $S^1$, $S^2$, $\Real_+$ and $\Real
$. We emphasize again that the curvature of the direct product manifold
$M$ is mainly due to the spherical components.

Consider a data set $x_1, \ldots, x_n \in M$, where $x_i \equiv
(x_i^1,\ldots, x_i^d)$ such that $x_i^j \in M_j$. Denote $d_0 \ge d$
for the intrinsic dimension of $M$. The geodesic mean $\bar{x}$ of the
data is defined component-wise for each simple manifold $M_j$.
Similarly, the tangent plane at $\bar{x}$, $T_{\bar{x}} M$, is also
defined marginally, that is, $T_{\bar{x}} M$ is a direct product of
tangent spaces of the simple manifolds. This tangent space gives a way
of applying Euclidean space-based statistical methods by mapping the
data onto $T_{\bar{x}} M$. We can manipulate this approximation of the
data component-wise. In particular, the marginal data on the $S^2$
components can be represented in a linear space by a transformation
$h_{\bar\delta}$, depending on the principal circles, that differs from
the tangent space approximation.

Since the principal circles $\bar\delta$ capture the nongeodesic
directions of variation, we use the principal circles as axes, which
can be thought of as flattening the quadratic form of variation. In
principle, we require a mapping $h_{\bar\delta}\dvtx  S^2 \rightarrow\Real
^2$ to have the following properties:
For $\bar\delta= (\delta_1, \delta_2) = ((\delta_1(\mathbf{c},r),\mathbf{u})$:
\begin{itemize}
\item$\mathbf{u}$ is mapped to the origin,
\item$\delta_1$ is mapped onto the $x$-axis, and
\item$\delta_2$ is mapped onto the $y$-axis.
\end{itemize}
Two reasonable choices of the mapping $h_{\bar\delta}$ will be
discussed in Section~\ref{sec:LinearRep} in detail.

The mapping $h_{\bar\delta}$ and the tangent space projection together
give a linear space representation of the data where the Euclidean PCA
is applicable. The line segments corresponding to the sample principal
component direction of the transformed data can be mapped back to $M$,
and become the principal arcs.

A procedure for principal arc analysis is as follows:
\begin{enumerate}[(2)]
\item[(1)] For each $j$ such that $M_j$ is $S^2$, compute principal
circles $\bar\delta= \bar\delta(x_1^j,\ldots, x_2^j)$ and the ratio
$\widehat{\mu/\sigma}$. If the ratio is greater than the predetermined
value $\varepsilon= 2$, then $\bar\delta$ is adjusted to be great circles
as explained in Section~\ref{PAAonSphere}.
\item[(2)] Let $\mathbf{h}\dvtx  M \rightarrow\Real^{d_0}$ be a
transformation $\mathbf{h}(x) = (h_1(x^1),\ldots,h_d(x^d))$. Each
component of $\mathbf{h}$ is defined as
\begin{eqnarray*}
h_j(x^j) =  \cases{
h_{\bar{\delta}} (x^j) &\quad\mbox{for }$M_j = {S^2}$, \cr
\Log_{\bar{x}^j} (x^j) &\quad\mbox{otherwise},
}
\end{eqnarray*}
where $\Log_{\bar{x}^j}$ and $h_{\bar{\delta}}$ are defined in the
\hyperref[append:manifold]{Appendix} and Section~\ref{sec:LinearRep}, respectively.
\item[(3)] Observe that $\mathbf{h}(x_1),\ldots,\mathbf{h}(x_n) \in
\Real^{d_0}$ always have their mean at the origin. Thus, the singular
value decomposition of the $d_0 \times n$ data matrix $\mathbf{X}
\equiv[\mathbf{h}(x_1) \cdots\mathbf{h}(x_n)]$ can be used for
computation of the PCA. Let $\mathbf{v}_1, \mathbf{v}_2, \ldots,\mathbf{v}_m$ be the left singular vectors of $\mathbf{X}$ corresponding to
the largest $m$ singular values.
\item[(4)] The $k$th principal arc is obtained by mapping the direction
vectors $\mathbf{v}_k$ onto $M$ by the inverse of $h$, which can be
computed component-wise.
\end{enumerate}

The principal arcs on $M$ are not, in general, geodesics. Nor are they
necessarily circles, in the original marginal $S^2$. This is because
$h_{\bar{\delta}}$ and its inverse $h_{\bar{\delta}}^{-1}$ are
nonlinear transformations and, thus, a line on $\Real^2$ may not be
mapped to a circle in $S^2$. This is consistent with the fact that the
principal components on a subset of variables are different from
projections of the principal components from the whole variables.

Principal arc analysis for data on direct product manifolds often
results in a concise summary of the data. When we observe a significant
variation along a small circle of a marginal $S^2$, that is most likely
not a random artifact but, instead, the result of a signal driving the
circular variation. Nongeodesic variation of this type is well captured
by our method.

Principal arcs can be used to reduce the intrinsic dimensionality of
$M$. Suppose we want to reduce the dimension by $k$, where $k$ can be
chosen by inspection of the scree plot. Then each data point $x$ is
projected to a $k$-dimensional submanifold $M_0$ of $M$ in such a way that
\[
\mathbf{h}^{-1} \Biggl( \sum_{i=1}^k \mathbf{v}_i \mathbf{v}_i' \mathbf{h}(x)  \Biggr) \in M_0,
\]
where the $\mathbf{v}_i$'s are the principal direction vectors in $\Real
^{d_0}$, found by step~3 above.
Moreover, the manifold $M_0$ can be parameterized by the $k$ principal
components $z_1, \ldots, z_k$ such that $M_0(z_1,\ldots,z_k) = \mathbf{h}^{-1}( \sum_{i=1}^k z_i \mathbf{v}_i)$.

\subsection{Choice of the transformation $h_{\bar\delta}$}\label{sec:LinearRep}

The transformation $h_{\bar\delta}\dvtx S^2 \rightarrow\Real^2$ leads to an
alternative representation of the data, which differs from the tangent
space projection. The $h_{\bar\delta}$ transforms nongeodesic scatters
along $\delta_1$ to scatters along the $x$-axis, which makes a linear
method like the PCA applicable. Among many choices of transformations
that satisfy the three principles we stated, two methods are discussed
here. Recall that $\bar\delta= (\delta_1,\delta_2) = (\delta_1(\mathbf{c},r), \mathbf{u})$.

\textit{Projection}. The first approach is based on the projection of
$\mathbf{x}$ onto $\delta_1$, defined in (\ref{eq:projectiondelta1}),
and a residual $\xi$. The signed distance from $\mathbf{u}$ to
$P_{\delta_1} x$, whose unsigned version is defined in (\ref{eq:intrinsinU}), becomes the $x$-coordinate, while the residual $\xi$
becomes the $y$-coordinate. This approach has the same spirit as the
model for the circle class (\ref{eq:modelForError}), since the
direction of the signal is mapped to the $x$-axis, with the perpendicular
axis for errors.

The projection $h_{\bar\delta} (\mathbf{x})$ that we define here is
closely related to the spherical coordinate system. Assume $\mathbf{c}
= \mathbf{e}_3$, and $\mathbf{u}$ is at the Prime meridian (i.e., on
the $x-z$ plane). For $\mathbf{x}$ and its spherical coordinates $(\phi
, \theta)$ such that $\mathbf{x}=(x_1, x_2, x_3) = (\cos\phi\sin\theta
, \cos\phi\sin\theta,\cos\theta)$,
%
\begin{equation}\label{eq:projectionh}
h_{\bar\delta} (\mathbf{x}) = \bigl(\sin(r) \phi, \theta- \theta_\mathbf{u} \bigr),
\end{equation}
where $\theta_\mathbf{u} = \cos^{-1}(u_3)$ is the latitude of $\mathbf{u}$. The set of $h_{\bar\delta}(\mathbf{x}_i)$ has mean zero because
the principal circle mean $\mathbf{u}$ has been subtracted.

\textit{Conformal map}.
A conformal map is a function which preserves angles. We point out two
conformal maps that can be combined to serve our purpose. See Chapter 9
of \citet{Churchill1984} and \citet{Krantz1999} for detailed discussions
of conformal maps. A~conformal map is usually defined in terms of
complex numbers. Denote the extended complex plane $\mathbb{C} \cup\{
\infty\}$ as $\mathbb{C}^*$. Let $\phi_\mathbf{c}: S^2 \rightarrow
\mathbb{C}^*$ be the stereographic projection of the unit sphere when
the point antipodal from $\mathbf{c}$ is the projection point. Then
$\phi_\mathbf{c}$ is a bijective conformal mapping defined for all
$S^2$ that maps $\delta_1$ as a circle centered at the origin in
$\mathbb{C}^*$.
The linear fractional transformation, sometimes called the M\"{o}bius
transformation, is a rational function of complex numbers, that can be
used to map a circle to a line in $\mathbb{C}^*$. In particular, we
define a linear fractional transformation $f_{u^*}\dvtx \mathbb{C}^*
\rightarrow\mathbb{C}^*$ as
%
\begin{equation}\label{eq:conformalh}
f_{u^*}(z) =  \cases{
\displaystyle \frac{\alpha i(z-u^*)}{-z - u^*} &\quad \mbox{if }$z \ne- u^*$, \cr
\infty&\quad \mbox{if }$z = - u^*$,
}
\end{equation}
where $u^* = \phi_\mathbf{c}(\mathbf{u})$, and $\alpha$ is a constant
scalar. Then the image of $\delta_1$ under $f_{u^*}\circ\phi_c$ is the
real axis, while the image of $\delta_2$ is the imaginary axis. The
mapping $h_{\bar\delta}\dvtx  S^2 \rightarrow\Real^2$ is defined by
$f_{u^*}\circ\phi_c$ with the resulting complex numbers understood as
members of $\Real^2$.
Note that orthogonality of any two curves in $S^2$ is preserved by the
$h_{\bar\delta}$ but the distances are not. Thus, we use the scale
parameter $\alpha$ of the function $f_{u^*}$ to match the resulting
total variance of $h_{\bar\delta}(\mathbf{x}_i)$ to the geodesic
variance of $\mathbf{x}_i$.

\begin{figure}

\includegraphics{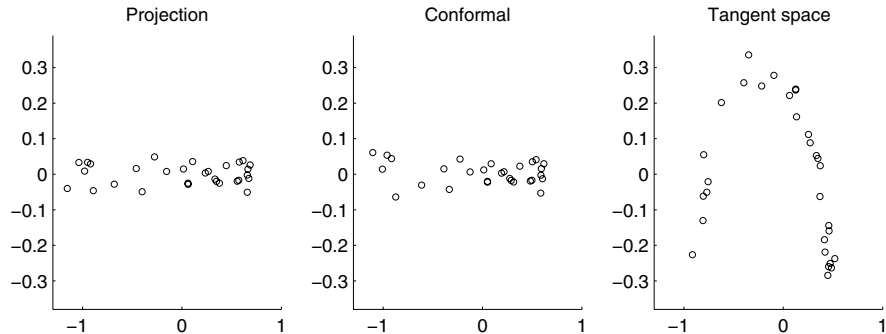}

\caption{Illustration of projection $h_{\bar\delta}$ [left, equation
(\protect\ref{eq:projectionh})] and conformal $h_{\bar\delta}$ [center,
equation (\protect\ref{eq:conformalh})] compared to a tangent plane projection
at the geodesic mean (right) of the data in Figure~\protect\ref{fig:smallCircleExamples}\textup{c}.
The $h_{\bar\delta}$ maps the variation along
$\delta_1$ to the variation along the $x$-axis, while the tangent plane
mapping fails to do so.}
\label{fig:LinRep}
\end{figure}

In many cases, both projection and conformal $h_{\bar\delta}$ give
better representations than just using the tangent space.
Figure~\ref{fig:LinRep} illustrates the image of $h_{\bar\delta}$ with
the toy data set depicted in Figure~\ref{fig:smallCircleExamples}c.
The tangent space mapping is also plotted for comparison.
The tangent space mapping leaves the curvy form of variation, while both
$h_{\bar\delta}$'s capture the variation and lead to an elliptical
distribution of the transformed data.

The choice between the projection and conformal mappings is a matter of
philosophy. The image of the projection $h_{\bar\delta}$ is not all of
$\Real^2$, while the image of the conformal $h_{\bar\delta}$ is all of
$\Real^2$. However, in order to cover $\Real^2$ completely, the
conformal $h_{\bar\delta}$ can grossly distort the covariance structure
of the data. In particular, the data points that are far from $\mathbf{u}$ are sometimes overly diffused when the conformal $h_{\bar\delta}$
is used, as can be seen in the left tail of the conformal mapped image
in Figure~\ref{fig:LinRep}. The projection $h_{\bar\delta}$ does not
suffer from this problem. Moreover, the interpretation of projection
$h_{\bar\delta}$ is closely related to the circle class model.
Therefore, we recommend the projection $h_{\bar\delta}$, which is used
in the following data analysis.

\section{Application to m-rep data}\label{sec:realData}
In this section an application of Principal Arc Analysis to the medial
representation (m-rep) data example, introduced below in more detail,
is described.

\subsection{Medial representation}
\begin{figure}

\includegraphics{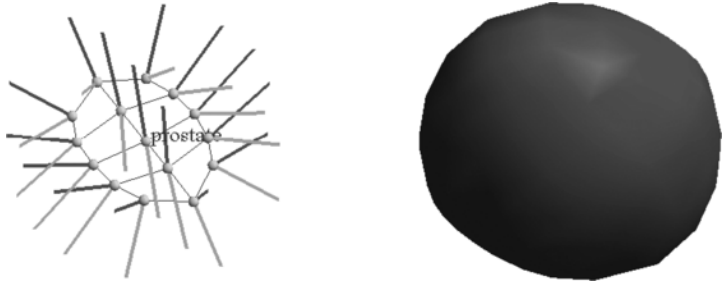}

\caption{(Left) An m-rep model with $3\times5$ grids of medial atoms.
Each atom has its location ($\Real^3$) and two equal-length spokes
($\Real_+ \otimes S^2 \otimes S^2$). (Right) The implied surface of the
m-rep model, showing a prostate shape. }
\label{fig:prostateFigure}
\end{figure}

The m-rep gives an efficient way of representing 2- or 3-dimensional
objects. The m-rep is constructed from the \textit{medial axis}, which
is a means of representing the middle of geometric objects. The {medial
axis} of a 3-dimensional object is formed by the centers of all spheres
that are interior to objects and tangent to the object boundary at two
or more points. In addition, the medial description is defined by the
centers of the inscribed spheres and by the associated vectors, called
spokes, from the sphere center to the two respective tangent points on
the object boundary. The medial axis is sampled over an approximately
regular lattice and the elements of the lattice are called \textit{medial atoms}. A~medial atom consists of the location of the atom
combined with two equal-length spokes, defined as a 4-tuple:
\begin{itemize}
\item location in $\Real^3$;
\item spoke direction 1, in $S^2$;
\item spoke direction 2, in $S^2$;
\item common spoke length in $\Real_+$;
\end{itemize}
as shown in Figure~\ref{fig:prostateFigure}. The size of the regular
lattice is fixed for each object in practice. For example, the shape of
a prostate is usually described by a $3 \times5$ grid of medial atoms,
across all samples. The collection of the medial atoms is called the
medial representation (m-rep). An m-rep corresponds to a particular
shape of prostate, and is a point in the m-rep space $\mathcal{M}$. The
space of prostate m-reps is then $\mathcal{M} = (\Real^3 \otimes\Real
_+ \otimes S^2 \otimes S^2 )^{15}$, which is a 120-dimensional direct
product manifold with $60$ components. The m-rep model provides a
useful framework for describing shape variability in intuitive terms.
See \citet{Siddiqi2008} and \citet{Pizer2003} for detailed introduction
to and discussion of this subject.

An important topic in medical imaging is developing segmentation
methods of 3D objects from CT images; see \citet{Cootes2001} and \citet{Pizer2007}. A~popular approach is similar to a Bayesian estimation
scheme, where the knowledge of anatomic geometries is used (as a prior)
together with a measure of how the segmentation matches the image (as a
likelihood). A~prior probability distribution is modeled using m-reps
as a means of measuring geometric atypicality of a segmented object.
PCA-like methods (including PAA) can be used to reduce the
dimensionality of such a model. A~detailed description can be found in
\citet{Pizer2007}.

\subsection{Simulated m-rep object}
The data set partly plotted in Figure~\ref{fig:prostateExample} is from
the generator discussed in \citet{Jeong2008}. It generates random
samples of objects whose shape changes and motions are physically
modeled (with some randomness) by anatomical knowledge of the bladder,
prostate and rectum in the male pelvis. Jeong et al. have proposed and
used the generator to estimate the probability distribution model of
shapes of human organs.

In the data set of 60 samples of prostate m-reps we studied, the major
motion of prostate is a rotation. In some $S^2$ components, the
variation corresponding to the rotation is along a small circle.
Therefore, PAA should fit better for this type of data than principal
geodesics. To make this advantage more clear, we also show results from
a data set by removing the location and the spoke length information
from the m-reps, the sample space of which is then $\{S^2\}^{30}$.

We have applied PAA as described in the previous section. The ratios
$\mu/\sigma$, estimated for the 30 $S^2$ components, are in general
large (with minimum 21.2, median 44.1 and maximum 118), which suggests
use of small circles to capture the variation.

\begin{figure}

\includegraphics{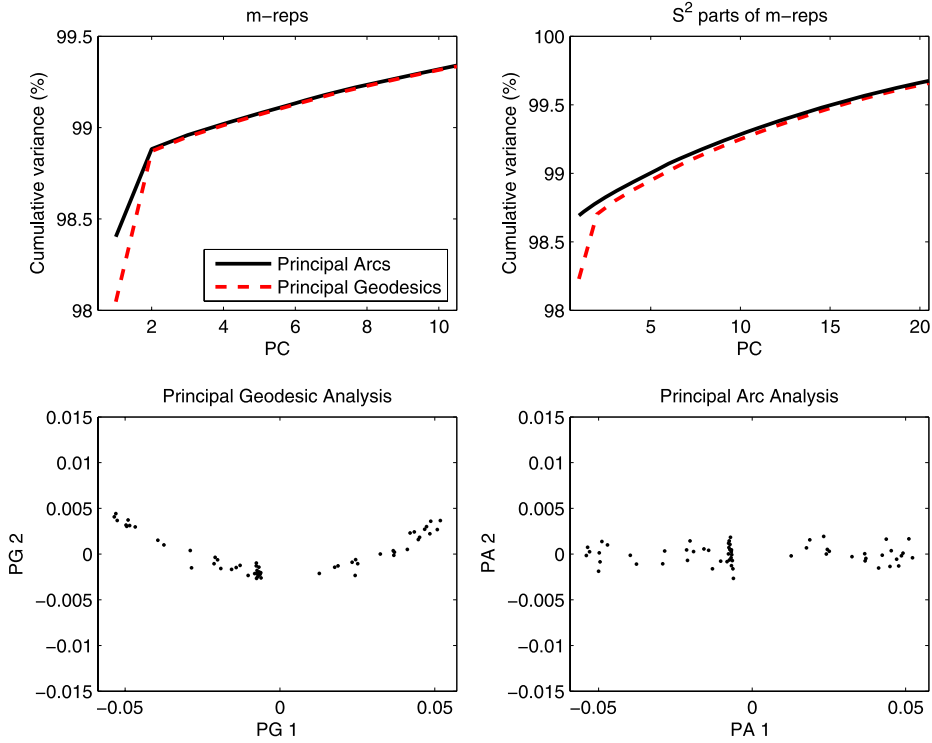}

\caption{(Top) The proportion of variances captured by the first few
components of PAA are compared to those from PGA for the simulated
prostate m-reps. (Bottom) Scatter plots of the data on $\{S^2\}^{30}$
show that the major variation is explained more concisely by the first
principal arc.}
\label{fig:prostatecumul}
\end{figure}

Figure~\ref{fig:prostatecumul} shows the proportion of the cumulative
variances, as a function of number of components, from the Principal
Geodesic Analysis (PGA) of \citet{Fletcher2004a} and PAA. In both cases,
the first principal arc leaves smaller residuals than the first
principal geodesic. What is more important is illustrated in the
scatterplots of the data projected onto the first two principal
components. The quadratic form of variation that requires two PGA
components is captured by a single PAA component.
%

%

The probability distribution model estimated by principal geodesics is
qualitatively different from the distribution estimated by PAA.
Although the difference in the proportion of variance captured is
small, the resulting distribution from PAA is no longer elliptical. In
this sense, PAA gives a convenient way to describe a nonelliptical
distribution by, for example, a Normal density.

\subsection{Prostate m-reps from real patients}
We also have applied PAA to a prostate m-rep data set from real CT
images. Our data consist of five patients' image sets, each of which is
a series of CT scans containing prostate taken during a series of
radiotherapy treatments [\citet{Merck2008}]. The prostate in each image
is manually segmented by experts and an m-rep model is fitted. The
patients, coded as 3106, 3107, 3109, 3112 and 3115, have different
numbers of CT scans (17, 12, 18, 16 and 15, respectively). We have in
total 78 m-reps.

\begin{figure}

\includegraphics{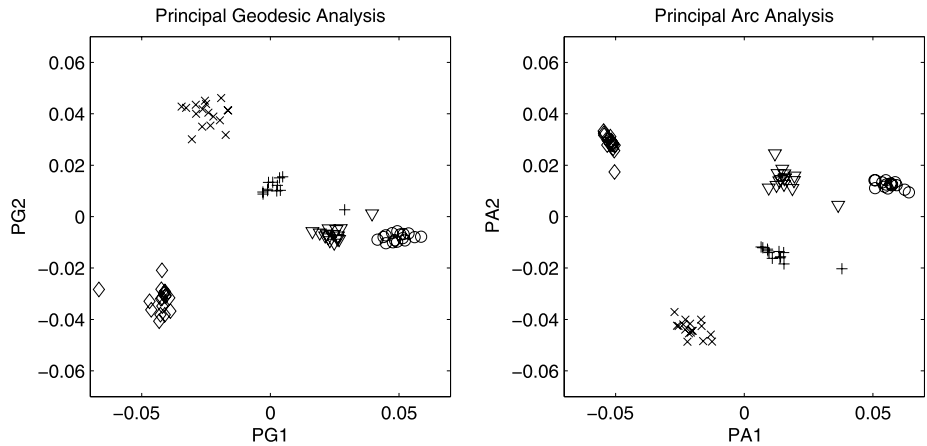}

\caption{The scatter plots of the real prostate m-rep data. Different
symbols represent different patients. PAA (right) gives a better
separation of different patients in the first two components compared
to PGA (left).}
\label{fig:realprostate}
\end{figure}

The proportion of variation captured in the first principal arc is
40.89\%, slightly higher than the 40.53\% of the first principal geodesic.
Also note that the estimated probability distribution model from PAA is
different from that of PGA.
In particular, PAA gives a better separation of patients in the first
two components, as depicted in the scatter plots (Figure~\ref{fig:realprostate}).

\section{Doubly iterative algorithm to find the least-squares small
circle}\label{sec:algorithm}

We propose an algorithm to fit the least-squares small circle (\ref{eq:getSmallCircle}),
which is a constrained nonlinear minimization
problem. This algorithm is best understood in two iterative steps: The
outer loop approximates the sphere by a tangent space; the inner loop
solves an optimization problem in the linear space, which is much
easier than solving (\ref{eq:getSmallCircle}) directly. In more detail,
the $(k+1)$th iteration works as follows. The sphere is approximated by a
tangent plane at $\mathbf{c}_k$, the $k$th solution of the center of
the small circle. For the points on the tangent plane, any iterative
algorithm to find a least-squares circle can be applied as an inner
loop. The solution of the inner iteration is mapped back to the sphere
and becomes the $(k+1)$th input of the outer loop operation.
One advantage of this algorithm lies in the reduced difficulty of the
optimization task. The inner loop problem is much simpler than (\ref{eq:getSmallCircle}) and the outer loop is calculated by a closed-form
equation, which leads to a stable and fast algorithm. Another advantage
can be obtained by using the \textit{exponential map} and \textit{log
map} (\ref{eq:ExpcLogc}) for the tangent projection, since they
preserve the distance from the point of tangency to the others, that
is, $\rho(\mathbf{x},\mathbf{c}) = \|\Log_\mathbf{c} (\mathbf{x})\|$
for any $\mathbf{x} \in S^2$. This is also true for radii of circles.
The exponential map transforms a circle in $\Real^2$ centered at the
origin with radius $r$ to $\delta(\mathbf{c},r)$. Thus, whenever (\ref{eq:getSmallCircle}) reaches its minimum, the algorithm does not alter
the solution.

We first illustrate necessary building blocks of the algorithm.
A~tangent plane $T_\mathbf{c}$ at $\mathbf{c}$ can be defined for any
$\mathbf{c}$ in $S^2$, and an appropriate coordinate system of
$T_\mathbf{c}$ is obtained as follows. Basically, any two orthogonal
complements of the direction $\mathbf{c}$ can be used as coordinates of
$T_\mathbf{c}$. For example, when $\mathbf{c} = (0,0,1)' \equiv\mathbf{e}_3$, a coordinate system is given by $\mathbf{e}_1$ and $\mathbf{e}_2$. For a general $\mathbf{c}$, let $q_\mathbf{c}$ be a rotation
operator on $\Real^3$ that maps $\mathbf{c}$ to $\mathbf{e}_3$. Then a
coordinate system for $T_\mathbf{c}$ is given by the inverse of
$q_\mathbf{c}$ applied to $\mathbf{e}_1$ and $\mathbf{e}_2$, which is
equivalent to applying $q_\mathbf{c}$ to each point of $S^2$ and using
$\mathbf{e}_1$, $\mathbf{e}_2$ as coordinates.

The rotation operator $q_\mathbf{c}$ can be represented by a rotation matrix.
For $\mathbf{c} = (c_x,c_y,c_z)'$, the rotation $q_\mathbf{c}$ is
equivalent to rotation through the angle $\theta= \cos^{-1} (c_z)$
about the axis
$\mathbf{u} = (c_y,-c_x,0)'/\sqrt{1-c_z^2}$, whenever $\mathbf{c} \ne
\pm\mathbf{e}_3$. When $\mathbf{c} = \pm\mathbf{e}_3$, $\mathbf{u}$ is
set to be $\mathbf{e}_1$. It is well known that a rotation matrix with
axis $\mathbf{u} = (u_x, u_y, u_z)'$ and angle $\theta$ in radians is,
for $c = \cos(\theta)$, $s = \sin(\theta)$ and $v = 1-\cos(\theta)$,
%
\begin{equation}
R_\mathbf{c} =  \pmatrix{
c + u_x^2 v & u_x u_y v - u_z s & u_x u_z v + u_y s \cr
u_x u_y v + u_z s & c+u_y^2 v & u_yu_zv-u_x s \cr
u_x u_z v - u_y s & u_y u_z v+u_x s & c+ u_z^2 v
}
,
\end{equation}
so that $q_\mathbf{c}(\mathbf{x}) = R_\mathbf{c} \mathbf{x} $, for
$\mathbf{x} \in\Real^3$.

With the coordinate system for $T_\mathbf{c}$, we shall define the
exponential map $\Exp_\mathbf{c}$, a mapping from $T_\mathbf{c}$ to
$S^2$, and the log map $\Log_\mathbf{c} = \Exp_\mathbf{c}^{-1}$. These
are defined for $\mathbf{v} = (v_1,v_2) \in\Real^2$ and $\mathbf{x} =
(x_1, x_2, x_3)' \in S^2$, as
%
\begin{equation}\label{eq:ExpcLogc}
\Exp_\mathbf{c} (\mathbf{v}) = q_\mathbf{c} \circ\Exp_{\mathbf{e}_3}(\mathbf{v}),\qquad
\Log_\mathbf{c} (\mathbf{x}) = \Log_{\mathbf{e}_3}\circ\, q_\mathbf{c} (\mathbf{x}),
\end{equation}
%
for $\theta= \cos^{-1} (x_3)$. See (\ref{eq:Exp_p})--(\ref{eq:Log_p}) in the \hyperref[append:manifold]{Appendix}  for $\Exp_{\mathbf{e}_3}$ and $\Log_{\mathbf{e}_3}$.
Note that $\Log_{\mathbf{c}}(\mathbf{c}) = \mathbf{0}$ and $\Log
_{\mathbf{c}}$ is not defined for the antipodal point of $\mathbf{c}$.

Once we have approximated each $\mathbf{x}_i$ by $\Log_\mathbf{c}
(\mathbf{x}_i) \equiv\tilde{\mathbf{x}}_i$, the inner loop finds the
minimizer $(\mathbf{v},r)$ of
%
\begin{equation}\label{eq:getCircle}
\min \sum_{i=1}^{n}  (\|\tilde{\mathbf{x}}_i- \mathbf{v}\| - r
 )^2,
\end{equation}
which is to find the least-squares circle centered at $\mathbf{v}$ with
radius $r$. The general circle fitting problem is discussed in, for
example, \citet{Umbach2003} and \citet{Chernov2010}. This problem is much
simpler than (\ref{eq:getSmallCircle}) because it is an unconstrained
problem and the number of parameters to optimize is decreased by 1.
Moreover, optimal solution of $r$ is easily found as
%
\begin{equation}\label{eq:getCircleRadius}
\hat{r} = \frac{1}{n}\sum_{i=1}^{n} \|\tilde{\mathbf{x}}_i - \mathbf{v}\|,
\end{equation}
when $\mathbf{v}$ is given. Note that for great circle fitting, we can
simply put $\hat{r} = \pi/2$. Although the problem is still nonlinear,
one can use any optimization method that solves nonlinear least squares
problems. We use the Levenberg--Marquardt algorithm, modified by \citet{Fletcher1971} [see Chapter 4 of \citet{Scales1985} and Chapter 3 of
\citet{Bates1988}], to minimize (\ref{eq:getCircle}) with $r$ replaced
by $\hat{r}$. One can always use $\mathbf{v} =\mathbf{0}$ as an initial
guess since $\mathbf{0} = \Log_\mathbf{c} (\mathbf{c})$ is the solution
from the previous (outer) iteration.

The algorithm is now summarized as follows:
\begin{enumerate}[(1)]
\item[(1)] Given $\{ \mathbf{x}_1 , \ldots,\mathbf{x}_n \}$, $\mathbf{c}_0 =
\mathbf{x}_1$.
\item[(2)] Given $\mathbf{c}_k$, find a minimizer $\mathbf{v}$ of (\ref{eq:getCircle}) with $r$ replaced by (\ref{eq:getCircleRadius}), with
inputs $\tilde{\mathbf{x}}_i = \Log_{\mathbf{c}_k} (\mathbf{x}_i)$.
\item[(3)] If $\norm{\mathbf{v}} < \varepsilon$, then iteration stops with the
solution $\hat{\mathbf{c}} = \mathbf{c}_{k}$, $r = \hat{r}$ as in (\ref{eq:getCircleRadius}). Otherwise, $\mathbf{c}_{k+1} = \Exp_{\mathbf{c}_k}(\mathbf{v})$ and go to step 2.
\end{enumerate}

Note that the radius of the fitted circle in $T_\mathbf{c}$ is the same
as the radius of the resulting small circle. There could be many
variations of this algorithm: as an instance, one can elaborate the
initial value selection by using the eigenvector of the sample
covariance matrix of $\mathbf{x}_i$'s, corresponding to the smallest
eigenvalue as done in \citet{Gray1980}. Experience has shown that the
proposed algorithm is stable and speedy enough. Gray, Geiser and Geiser
proposed to solve (\ref{eq:getSmallCircle}) directly, which seems to be
unstable in some cases.

The idea of the doubly iterative algorithm can be applied to other
optimization problems on manifolds. For example, the geodesic mean is
also a solution of a nonlinear minimization, where the nonlinearity
comes from the use of the geodesic distance. This can be easily solved
by an iterative approximation of the manifold to a linear space [See
Chapter 4 of \citet{Fletcher2004}], which is the same as the gradient
descent algorithms [\citet{Pennec1999}, \citet{Le2001}]. Note that the
proposed algorithm, like other iterative algorithms, only finds one
solution even if there are multiple solutions.

\begin{appendix}
\section*{Appendix: Some background on direct product manifold}\label{append:manifold}

We give some necessary geometric background on direct product
manifolds. Precise definitions and geometric discussions on a richer
class of manifold, Riemannian manifold, can be found in \citet{Boothby1986} and \citet{Helgason2001}.

A $d$-dimensional manifold can be thought of as a curved surface
embedded in a Euclidean space of higher dimension $d'$ (${\ge}d$). The
manifold is required to be smooth, that is, infinitely differentiable,
so that a sufficiently small neighborhood of any point on the manifold
can be well approximated by a linear space. The \textit{tangent space} at
a point $p$ of a manifold $M$, $T_pM$, is defined as a linear space of
dimension $d$ which is tangent to $M$ at $p$. The notion of distance on
$M$ is handled by a Riemannian metric, which is a metric of tangent
spaces. In particular, the \textit{geodesic distance function} $\rho
_M(p,q)$ is roughly defined as the length of the shortest curve joining
$p,q \in M$.

Now consider a direct product manifold $M = M_1 \otimes\cdots\otimes
M_{m}$. We restrict our attention to the case where each $M_i$ is one
of the simple manifolds $S^1, S^2, \Real_+$ or $\Real$, but most of the
assertions below apply equally well to direct products of more general
manifolds.

\textit{Geodesic distance function}. The geodesic distance between $p
\equiv(p^1,\ldots, p^{m})$ and $q \equiv(q^1,\ldots, q^{m})$ is
defined by
\[
\rho_M(p,q) =  \Biggl(\sum_{i=1}^{m} \rho_{M_i}^2 (p^i, q^i)  \Biggr)^{1/2},
\]
where each $\rho_{M_i}$ is the geodesic distance function on $M_i$. The
geodesic distance on $S^1$ is defined by the length of the shortest
arc. Similarly, the geodesic distance on $S^2$ is defined by the length
of the shortest great circle segment. The geodesic distance on $\Real
_+$ needs special treatment. In many practical applications, $\Real_+$
represents a space of scale parameters. A~desirable property for a
metric on scale parameters is \textit{scale invariance}, $\rho(rx,ry) =
\rho(x,y)$ for any $x,y,r \in\Real_+$. This can be achieved by
differencing the logs, that is,
%
\begin{equation}\label{eq:metric_Real+}
\rho_{\Real_+}(x,y) = \bigg|\log\frac{x}{y} \bigg|\qquad \mbox{for } x,y \in
\Real_+.
\end{equation}
Finally, the geodesic distance on a simple manifold $\Real$ or $\Real
^d$ is the Euclidean distance.

\textit{Geodesic mean and variance}. The \textit{geodesic mean} of a set
of points in $M$, also referred to as the intrinsic mean, is also
calculated component-wise. The \textit{geodesic mean} of $x_1,\ldots, x_n
\in M$ is the minimizer in $M$ of the sum of squared geodesic distances
to the data. Thus, the geodesic mean is defined as
%
\begin{equation}\label{eq:geodesic_mean}
\bar{x}= \argmin_{x\in M} \frac{1}{n}\sum_{i=1}^n \rho_M^2(x,x_i).
\end{equation}
In fact, each $\bar{x}^i$ of $\bar{x} = (\bar{x}^1, \ldots, \bar{x}^m)$
is the geodesic mean of $x_1^i,\ldots, x_n^i \in M_i$. The geodesic
mean of $\theta_1, \ldots, \theta_n \in[0,2\pi) \cong S^1$ is found by
examining a candidate set consisting of
%
\begin{equation}\label{eq:geodmeanS1}
\bar\theta_j = \frac{\sum_{i=1}^{n}\theta_i + 2j\pi}{n},\qquad   j =
0,\ldots, n-1,
\end{equation}
as in \citet{Moakher2002}.
The geodesic mean in $S^2$ can be calculated by a full grid search or
an iterative algorithm, as described in Section~\ref{sec:algorithm}.
The geodesic mean in $\Real_+$ or $\Real$ is straightforward. Note that
the geodesic mean may not be unique. However, throughout the paper, we
have assumed that the data have a unique geodesic mean which is true in
most data analytic situations.
Statistical investigation of the geodesic mean on manifolds can be
found, for example, in Bhattacharya and Patrangenaru (\citeyear{Bhattacharya2003,Bhattacharya2005}) and \citet{Le2000}.

A related notion is \textit{geodesic variance}. A~sample geodesic
variance is defined by the average squared geodesic distances to the
geodesic mean, that is, $\frac{1}{n} \sum_{i=1}^{n} \rho^2_M (\bar
{x},x_i)$. When $M$ is indeed the Euclidean space, the geodesic
variance is the same as the total variance (the trace of the
variance--covariance matrix).

\textit{Mappings to tangent space}. The \textit{exponential map} maps a
point in $T_pM$ to $M$. The \textit{log map} is the inverse exponential
map whose domain is in $M$. For a direct product manifold $M$, the
mappings are also defined component-wise. For $S^1$, let $\theta\in
\Real$ denote an element of $T_\mathbf{p}S^1$ where $\mathbf{p}$ is set
to be $(1,0) \in S^1$ embedded in $\Real^2$. Then the exponential map
is defined as
\[
\Exp_\mathbf{p}(\theta) = (\cos\theta, \sin\theta).
\]
The corresponding log map of $\mathbf{x} = (x_1, x_2)$ is defined as
$\Log_\mathbf{p}(x) = \operatorname{sign}(x_2)\cdot\mathop{\mathrm{arccos}}(x_1)$.
For $S^2$, let $\mathbf{v} = (v_1, v_2)$ denote a tangent vector in
$T_\mathbf{p}S^2$. Let $\mathbf{p} = \mathbf{e}_3$, then the
exponential map $\Exp_\mathbf{p}\dvtx  T_\mathbf{p}S^2 \To S^2$ is defined by
%
\begin{equation}\label{eq:Exp_p}
\Exp_\mathbf{p} (\mathbf{v}) =  \biggl(\frac{v_1}{\norm{\mathbf{v}}}\sin
{\norm{\mathbf{v}}}, \frac{v_2}{\norm{\mathbf{v}}}\sin{\norm{\mathbf{v}}}, \cos{\norm{\mathbf{v}}} \biggr).
\end{equation}
This equation can be understood as a rotation of the base point $\mathbf{p}$ to the direction of $\mathbf{v}$ with angle $\norm{\mathbf{v}}$.
The corresponding log map for a point $\mathbf{x} = (x_1, x_2, x_3) \in
S^2$ is given by
%
\begin{equation}\label{eq:Log_p}
\Log_\mathbf{p}(x) =  \biggl( x_1 \frac{\theta}{\sin{\theta}}, x_2 \frac
{\theta}{\sin{\theta}}  \biggr),
\end{equation}
where $\theta= \mathop{\mathrm{arccos}} (x_3)$ is the geodesic distance from
$\mathbf{p}$ to $x$. Note that the antipodal point of $p$ is not in the
domain of the log map, that is, the domain of $\Log_\mathbf{p}$ is
$S^2/\{-\mathbf{p}\}$. In both $S^1$ and $S^2$, $\mathbf{p}$ can be set
to be any point on $S^1$ or $S^2$. The exponential and log map for an
arbitrary $T_\mathbf{p}$ can be defined together with a rotation
operator, which is defined in Section~\ref{sec:algorithm} for the case
of $S^2$.
The exponential map of $\Real_+$ is defined by the standard real
exponential function. The domain of the inverse exponential map, the
log map, is $\Real_+$ itself. Finally, the exponential map on $\Real$
is the identity map.
\end{appendix}


\printaddresses

\end{document}